\documentclass[12pt,dvips]{article}

\usepackage{rotating}
\usepackage{axodraw}
\usepackage{epsfig}
\usepackage{amssymb}
\usepackage{color}

\newcommand{\imag}{\Im {\rm m}}
\newcommand{\real}{\Re {\rm e}}
\newcommand{\s}{\\ \vspace*{-3.5mm}}

\def\lsim{\mathrel{\raise.3ex\hbox{$<$\kern-.75em\lower1ex\hbox{$\sim$}}}}
\def\gsim{\mathrel{\raise.3ex\hbox{$>$\kern-.75em\lower1ex\hbox{$\sim$}}}}

\begin{document}

\begin{titlepage}

\begin{flushright}
\end{flushright}

\vskip 1.5cm

\begin{center}
{\Large\bf Analysis of the Neutralino System in \\[1mm]
           Three--Body Leptonic Decays of Neutralinos}\\[1.5cm]
           {S.Y. Choi$^1$, B.C. Chung$^2$, J. Kalinowski$^3$, Y.G. Kim$^4$
           and K. Rolbiecki$^3$}
\end{center}

\vskip 0.5cm

{\small
\begin{center}
$^1$ {\it Department of Physics, Chonbuk National University,
               Jeonju 561-756, Korea}\\
$^2$ {\it Department of Physics, KAIST, Daejon 305--017, Korea} \\
$^3$ {\it Institute of Theor. Phys., Warsaw University, Hoza 69, 00681 Warsaw,
          Poland}\\
$^4$ {\it Department of Physics, Korea University, Seoul 136--701, Korea}
\end{center}
}

\vskip 4.cm
\begin{abstract}
\noindent
Neutralinos $\tilde{\chi}^0$ in supersymmetric theories, the spin--1/2
Majorana--type superpartners of the U(1) and SU(2) neutral electroweak gauge
bosons
and SU(2) neutral Higgs bosons, are expected to be among light supersymmetric
particles so that they  can be produced copiously
via direct pair
production and/or from cascade decays of other sparticles such as sleptons
at the planned Large Hadron Collider and the prospective International
Linear Collider. Considering the  prospects  of having  both  highly
polarized neutralinos and possibility of
reconstructing their decay rest frames, we provide a systematic
investigation of the three--body leptonic decays of the
neutralinos in the minimal supersymmetric standard model and demonstrate 
alternative ways for probing the
Majorana
nature of the neutralinos and CP violation in the neutralino system.
\end{abstract}
\end{titlepage}

\section{Introduction}
\label{sec:introduction}

The search for supersymmetry (SUSY) is one of the main goals at
present and future colliders  since SUSY is generally accepted as
one of the most promising concepts for physics beyond the standard
model (SM) \cite{MSSM}. All SUSY theories contain neutralinos, the
spin--1/2 Majorana superpartners of neutral gauge bosons and Higgs
bosons, that are expected to be among the light supersymmetric
particles that can be produced copiously at future high energy
colliders.\s

Once neutralino candidates have been detected at high energy colliders
such as the Large Hadron Collider (LHC) \cite{LHC} and the
International Linear Collider (ILC) \cite{ILC}, it is of great
importance to verify that the observed states are indeed the spin--1/2
superpartners of the neutral SM gauge and Higgs bosons.  For that
purpose, it will be crucial to measure their quantum numbers and to
confirm that they are indeed Majorana fermions
\cite{Bilenky,Moortgat-Pick1,SYChoi1,SYChoi2}.  Moreover, masses,
mixings, couplings and CP violating phases must be measured in a
model--independent way \cite{Neut1,Neut2,Nojiri} to reconstruct the
fundamental SUSY parameters and to verify the SUSY relations at the
electroweak scale, leading to a reliable extrapolation to the
grand unification scale or the Planck scale \cite{RGE}.\s

In this report we focus on probing the Majorana nature and CP
properties of neutralinos in the minimal
supersymmetric standard model (MSSM) through the {\it charge (C)
self--conjugate} three--body decays of polarized neutralinos into the
lightest neutralino $\tilde{\chi}^0_1$ and a lepton pair
$\ell^+\ell^-$:
\begin{eqnarray}
\tilde{\chi}^0_i \rightarrow\tilde{\chi}^0_1 \,\, \ell^+\, \ell^-
\label{eq:chain}
\end{eqnarray}
with $\ell=e$ or $\mu$ whose four--momenta can be measured with
great precision.  In particular,
the decays of the second lightest neutralino $\tilde{\chi}^0_2$
will be studied in more detail  since it is expected in most
supersymmetric scenarios \cite{SPS} that the three--body decay
mode has a significant branching fraction only for the second
lightest neutralino, while the  heavier neutralinos decay
mainly through two--body decays.

Thorough analysis of the CP properties and Majorana
nature of neutralinos produced in pairs in $e^+e^-$ annihilation   
has been performed in
Refs.~\cite{Moortgat-Pick1} and \cite{Neut2}  
[CP asymmetries in neutralino production with two-body decays have
been investigated in Ref.~\cite{twobody}].
The spin--1/2 neutralinos $\tilde{\chi}^0_i$  
are produced polarized with
the degree of polarization depending on their production
mechanism and polarization of the colliding beams.  
Because in  general the momenta of final particles do depend on the neutralino
polarization,  full account of spin correlations
between production and decay processes is
necessary~\cite{Moortgat-Pick1,Neut2}.  

Since the verification of Majorana character of neutralinos and their 
CP properties are of
fundamental importance, they have to be scrutinized in all possible ways. Here
we consider neutralinos $\tilde{\chi}^0_2$ which 
themselves are decay products of
scalar particles, {\it i.e.} sleptons.  In such a case 
their subsequent decay can be analyzed {\it independently} of
the production mechanism making the theoretical treatment greatly
simplified and the interpretation of various observables more transparent. 

As noted explicitly in a  
recent work \cite{Aguilar1}, neutralinos $\tilde{\chi}^0_2$
produced in $\tilde{e}^\pm_L$ decays are 100\% polarized, having
negative helicity in $\tilde{e}^-_L\to e^-\tilde{\chi}^0_2$ and
positive helicity in $\tilde{e}^+_L\to e^+\tilde{\chi}^0_2$.
Furthermore, it is possible to reconstruct the rest frame of the
neutralino $\tilde{\chi}^0_2$ in a few specific cascade processes,
for example, in the process 
\begin{eqnarray}
e^+e^-\to\tilde{e}^+_L\tilde{e}^-_L
\to e^+\tilde{\chi}^0_1 e^-\tilde{\chi}^0_2
\label{cascade}
\end{eqnarray}
followed by the
three--body decay $\tilde{\chi}^0_2\to \tilde{\chi}^0_1\mu^+\mu^-$
as shown in Refs.~\cite{Aguilar2} and \cite{Aguilar3}.
It is the purpose of this paper to show that  such a perfect
neutralino polarization combined with the study of angular correlations
in the neutralino rest frame  can provide us with alternative 
ways for  probing the Majorana nature of the
neutralinos\footnote{A clear independent evidence of the Majorana
character of the neutralinos can be provided by an experimental
identification of the selectron pair production in $e^-e^-$
collisions, which occur only via $t$-- and $u$--channel neutralino
exchange \cite{Aguilar2,e-e-}.} and CP violation in the neutralino
system. \s

Keeping in mind the above aspects,
we provide  in the present work a systematic
analysis of the neutralino decay $\tilde{\chi}^0_2 \to\tilde{\chi}^0_1
\ell^+\ell^-$ in  {\it its decay rest frame}
for extracting all the physical implications
due to the Majorana nature as well as CP
violation \cite{Moortgat-Pick1,Neut2,Aguilar2,Triple1,Triple2} in the
neutralino system of the MSSM   assuming  100\%  neutralino polarization.
Through the present work it is  assumed that the neutralino masses
have already been measured with great precision \cite{Baer2}. 
On the other hand, 
the efficiency of reconstructing the $\tilde{\chi}^0_2$ polarization as well
as its rest frame depends not only
on the cascade processes under consideration and on the values of relevant
SUSY parameters, but also on details of experimental setup. 
Experimental simulations with
realistic reconstruction efficiencies and with background processes
included, however, are beyond the scope of the semi-theoretical treatment  
in the present paper. Nevertheless, we hope that our findings are interesting
enough to motivate realistic simulations. \s 

The layout of the paper is as follows. In Sect.~\ref{sec:mixing}, we briefly 
recall the
mixing formalism for the neutral gauginos and higgsinos in the CP
non--invariant
theories with complex phases. In Section~\ref{sec:three_body_decay}
the leptonic decays $\tilde{\chi}^0_i\to \tilde{\chi}^0_1\ell^+\ell^-$
of a polarized neutralino $\tilde{\chi}^0_i$ ($i=2,3,4$) are described in 
terms of quartic
charges and Mandelstam kinematic variables, the
decay distribution in terms of two lepton energies and three angles in
the rest frame of the decaying neutralino is discussed. The
consequences of the CP and CP$\tilde{\rm T}$ invariance on the
polarized decay distributions are explained and new relations among the decay
amplitudes, {\it unique} for Majorana particles, are derived.  We illustrate in
Sect.~\ref{sec:majorana_cp} how to probe the Majorana character and CP
violation of the neutralino system through detailed analytical and
numerical investigation of various observables
in the  rest frame of the decaying neutralino $\tilde{\chi}^0_2$: CP--even 
lepton
energy/angular distributions, lepton invariant mass and opening angle
distributions, and a CP--odd triple product of the neutralino spin
vector and two lepton momenta. Finally, we summarize our findings and
conclude in Sect.~\ref{sec:conclusion}.\s

\section{Neutralino mixing}
\label{sec:mixing}

In the MSSM, the four neutralinos $\tilde{\chi}^0_i$ ($i=1,2,3,4$) are
mixtures of the neutral U(1) and SU(2) gauginos, $\tilde{B}$ and
$\tilde{W}^3$, and the SU(2) higgsinos, $\tilde{H}^0_1$ and
$\tilde{H}^0_2$. The neutralino mass matrix in the $(\tilde{B},
\tilde{W}^3, \tilde{H}^0_1, \tilde{H}^0_2)$ basis,
\begin{eqnarray}
{\cal M} =\left(\begin{array}{cccc}
          M_1  &  0  & -m_Z c_\beta s_W   &  m_Z s_\beta s_W  \\[1mm]
          0    & M_2 & m_Z c_\beta c_W    & -m_Z s_\beta c_W  \\[1mm]
       -m_Z c_\beta s_W &  m_Z c_\beta c_W &   0  & -\mu \\[1mm]
        m_Z s_\beta s_W & -m_Z s_\beta c_W & -\mu &  0
               \end{array}\right)
\label{eq:mass_matrix}
\end{eqnarray}
is built up by the fundamental SUSY parameters: the U(1) and SU(2)
gaugino masses $M_1$ and $M_2$, the higgsino mass parameter $\mu$, and
the ratio $\tan\beta=v_2/v_1$ of the vacuum expectation values of the
two neutral Higgs fields which break the electroweak symmetry. The
existence of CP--violating phases in supersymmetric theories in
general induces electric dipole moments (EDM). The current
experimental bounds on the EDM's can be exploited to derive indirect
limits on the parameter space \cite{CDG}.\s

By reparametrization of the fields, $M_2$ can be taken real and
positive without loss of generality so that the two remaining
non--trivial phases, which are reparametrization--invariant, may be
attributed to $M_1$ and $\mu$:
\begin{eqnarray}
M_1 = |M_1|\, {\rm e}^{i\Phi_1},\quad
\mu = |\mu|\, {\rm e}^{i\Phi_\mu}\ \ (0\leq \Phi_1, \Phi_\mu < 2\pi)
\end{eqnarray}
Since the  matrix ${\cal M}$ is symmetric, one unitary matrix $N$ is
sufficient to rotate the gauge eigenstate basis $(\tilde{B}, \tilde{W}^3,
\tilde{H}^0_1, \tilde{H}^0_2)$ to the mass eigenstate basis of the Majorana
fields $\tilde{\chi}^0_i$
\begin{eqnarray}
{\cal M}_{\rm diag} = N^*{\cal M} N^\dagger
\end{eqnarray}
The mass eigenvalues $m_i$ ($i=1,2,3,4$) in ${\cal M}_{\rm diag}$ can
be chosen real
and positive by a suitable definition of the unitary matrix
$N$.\s

The most general $4\times 4$ unitary matrix $N$ can be parameterized
by six angles and ten phases. It is convenient to factorize the matrix
$N$ into a diagonal Majorana--type $\mathbf{M}$ and a Dirac--type
$\mathbf{D}$ component in the following way:
\begin{eqnarray}
N=\mathbf{M}\mathbf{D}
\end{eqnarray}
The matrix $\mathbf{D}$ can be written as a sequence of 6 independent
two--dimensional rotations parameterized in terms of 6 angles and 6 phases.
The diagonal matrix $\mathbf{M}={\rm diag}\left\{{\rm e}^{i\alpha_1}, {\rm
    e}^{i\alpha_2}, {\rm e}^{i\alpha_3}, {\rm e}^{i\alpha_4}\right\}$ is given
in terms of Majorana phases $\alpha_i$ ($0\leq \alpha_i < \pi$) \cite{Neut1}.
One overall Majorana phase is non-physical and, for example, $\alpha_1$ may be
chosen to vanish.  \s

Due to the Majorana nature of the neutralinos, all nine phases of
the mixing matrix $N$ are fixed by underlying SUSY parameters, and
they cannot be removed by rephasing the fields. CP is conserved if
all the Dirac phases are 0  mod $\pi$ and the Majorana phases
$\alpha_i$ 0 mod $\pi/2$.  [Majorana phases $\alpha_i=\pm \pi/2$
describe different CP parities of the neutralino states.] As a
consequence, all the matrix elements $N_{i\alpha}$ are purely real
or purely imaginary in the CP invariant case.\s

\begin{figure}[!htb]
\begin{center}
\begin{picture}(350,160)(0,40)
\Text(0,120)[]{$\tilde{\chi}^0_i$}
\Line(10,120)(40,120)
\Line(40,120)(70,150)
\Text(79,155)[]{$\tilde{\chi}^0_1$}
\Photon(40,120)(60,90){2}{8}
\Text(40,98)[]{\color{blue} $Z$}
\ArrowLine(60,90)(90,90)
\Text(100,90)[]{$\ell^{-}$}
\ArrowLine(80,60)(60,90)
\Text(85,55)[]{$\ell^+$}
\Text(125,120)[]{$\tilde{\chi}^0_i$}
\Line(135,120)(165,120)
\ArrowLine(165,120)(195,150)
\Text(205,155)[]{$\ell^-$}
\DashArrowLine(185,90)(165,120){2}
\Text(165,100)[]{\color{blue} $\tilde{\ell}_{L,R}$}
\Line(185,90)(215,90)
\Text(225,90)[]{$\tilde{\chi}^0_1$}
\ArrowLine(205,60)(185,90)
\Text(210,55)[]{$\ell^+$}
\Text(250,120)[]{$\tilde{\chi}^0_i$}
\Line(260,120)(290,120)
\ArrowLine(320,150)(290,120)
\Text(330,155)[]{$\ell^+$}
\DashArrowLine(290,120)(310,90){2}
\Text(290,100)[]{\color{blue} $\tilde{\ell}_{L,R}$}
\Line(310,90)(340,90)
\Text(350,90)[]{$\tilde{\chi}^0_1$}
\ArrowLine(310,90)(335,55)
\Text(345,55)[]{$\ell^-$}
\end{picture}
\end{center}
\caption{\it Diagrams contributing to the leptonic three--body
             neutralino decay
             $\tilde{\chi}^0_i\to\tilde{\chi}^0_1\ell^+\ell^-$; the
             exchange of the neutral Higgs bosons is neglected
             because the contribution is strongly suppressed by the
             tiny electron and muon Yukawa couplings.}
\label{fig:diagrams}
\end{figure}
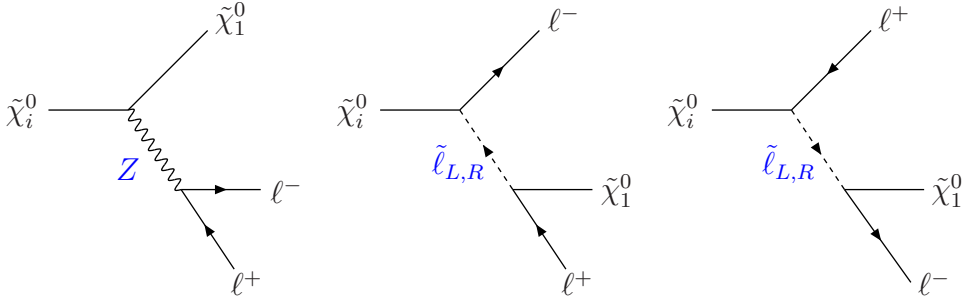
%

\section{Three--body leptonic neutralino decays}
\label{sec:three_body_decay}
%
\subsection{Neutralino decay amplitude}

The diagrams contributing to the three--body leptonic
decay process $\tilde{\chi}^0_i\rightarrow \tilde{\chi}^0_1\ell^+\ell^-$ are
shown in Fig.~\ref{fig:diagrams}. Here, the exchange of the neutral Higgs boson
[replacing the $Z$ boson] is neglected since the couplings to the
first and second generation SM leptons, $\ell=e$ and $\mu$, are very small.
In this case, all the components of the decay matrix elements are of the
(vector--current)$\times$(vector--current) form which, after a simple
Fierz transformation of the slepton--exchange parts, may be written
for the lepton
final states as
\begin{eqnarray}
D\left(\tilde{\chi}^0_i\rightarrow\tilde{\chi}^0_1
\ell^+\ell^-\right)
 = \frac{e^2}{m^2_{\tilde{\chi}^0_i}}D_{\alpha\beta}
   \left[\bar{u}(\tilde{\chi}^0_1) \gamma^\mu P_\alpha
               u(\tilde{\chi}^0_i)\right]
   \left[\bar{u}(\ell^-)  \gamma_\mu P_\beta  v(\ell^+)\right]
\label{eq:neutralino decay amplitude}
\end{eqnarray}
with the generalized bilinear charges $D_{\alpha\beta}$ ($\alpha,
\beta=L,R$) for the decay amplitudes:
\begin{eqnarray}
D_{LL}&=&+\frac{D_Z}{s_W^2c_W^2}(s_W^2 -\frac{1}{2}){\cal Z}_{1i}
              -D_{uL}\,g_{L1i}\nonumber\\
D_{LR}&=&+\frac{D_Z}{c_W^2}{\cal Z}_{1i}
              +D_{tR}\,g_{R1i}\nonumber\\
D_{RL}&=&-\frac{D_Z}{s_W^2c_W^2}(s_W^2 -\frac{1}{2}){\cal Z}^*_{1i}
              +D_{tL}\,g^*_{L1i}\nonumber\\
D_{RR}&=&-\frac{D_Z}{c_W^2}{\cal Z}^*_{1i}
              -D_{uR}\,g^*_{R1i}
\label{eq:bilinear_charges}
\end{eqnarray}
where the $s$--channel $Z$--boson, and the $t$-- and $u$--channel
slepton propagators are given by
\begin{eqnarray}
&& D_Z     =\frac{m^2_{\tilde{\chi}^0_i}}{s-m^2_Z+im_Z\Gamma_Z}\nonumber\\
&& D_{tL,R}=\frac{m^2_{\tilde{\chi}^0_i}}{t-m^2_{\tilde{e}_{L,R}}
                 +i m_{\tilde{e}_{L,R}} \Gamma_{\tilde{e}_{L,R}}}\nonumber\\
&& D_{uL,R}=\frac{m^2_{\tilde{\chi}^0_i}}{u-m^2_{\tilde{e}_{L,R}}
                 +i m_{\tilde{e}_{L,R}} \Gamma_{\tilde{e}_{L,R}}}
\end{eqnarray}
in terms of the Mandelstam variables, $s=(q_- + q_+)^2,\,
t=(p_i-q_-)^2$, and $u=(p_i-q_+)^2$, where $p_i$, $q_+$ and $q_-$ are
the 4--momenta of the decaying neutralino $\tilde{\chi}^0_i$ and the
positively and negatively charged leptons $\ell^\pm$,
respectively. The couplings ${\cal Z}_{ij}$, $g_{Lij}$ and $g_{Rij}$ are
given in terms of the neutralino diagonalization matrix elements
$N_{i\alpha}$ ($i,\alpha=1$ -- $4$):
\begin{eqnarray}
&& {\cal Z}_{ij}=\frac{1}{2}
                 \left(N_{i3}N^*_{j3}-N_{i4}N^*_{j4}\right)\,,\nonumber\\
&& g_{Lij}=\frac{1}{4 s_W^2c_W^2}(N_{i2}c_W+N_{i1}s_W)(N^*_{j2}c_W+N^*_{j1}s_W)
                 \nonumber\\
&& g_{Rij}=\frac{1}{c_W^2}N_{i1}N^*_{j1}
\end{eqnarray}
The complex couplings satisfy the hermiticity relations:
\begin{eqnarray}
{\cal Z}_{ij} = {\cal Z}^*_{ji},\quad
g_{Lij} = g^*_{Lji},\quad
g_{Rij} = g^*_{Rji}
\end{eqnarray}
so that, if the $Z$--boson and selectron widths are neglected in the $Z$ and
selectron propagators, all the bilinear charges $D_{\alpha\beta}$ also satisfy
similar relations.  In the CP invariant case the mixing matrix elements
$N_{i\alpha}$ are purely real or purely imaginary implying that the couplings
${\cal Z}_{ij}$, $g_{Lij}$ and $g_{Rij}$ are also purely real or purely
imaginary. However, these couplings are in general complex in the CP
non-invariant case, having both non--trivial real and imaginary parts.\s

\subsection{Neutralino decay distribution}

The absolute amplitude squared of the three--body leptonic decay
$\tilde{\chi}^0_i\to \tilde{\chi}^0_1 \ell^+\ell^-$ of a neutralino
$\tilde{\chi}^0_i$ with its polarization vector $n$ is given by
\cite{Triple1} (the full spin-density decay matrix  can be found in 
Refs.~\cite{Neut2,GMP})
\begin{eqnarray}
|{\cal D}|^2(n)&=& 4\,(m^2_i-t)\,(t-m^2_1)(N_1-N_3)
                  +4\,(m^2_i-u)(u-m^2_1)(N_1+N_3)\nonumber\\
                && -8\, m_i m_1\,s\,N_2
                   +16\, m_1\langle p_i n q_- q_+\rangle\, N_4 \nonumber\\
                && +8\, (n\cdot q_+)
                    \left[m_i (u-m^2_1)(N'_1+N'_3)-m_1(m^2_i-t)N'_2\right]
                    \nonumber\\
                && +8\, (n\cdot q_-)
                    \left[m_i (t-m^2_1)\,(N'_1-N'_3)+m_1 (m^2_i-u)N'_2\right]
\label{eq:decay distribution}
\end{eqnarray}
where $n$ is the $\tilde \chi^0_i$ spin 4-vector and $\langle
p_i n q_- q_+\rangle\equiv\epsilon_{\mu\nu\rho\sigma}
p_i^{\mu}n^{\nu}q_-^{\rho} q_+^{\sigma}$ with the convention
$\epsilon_{0123}=+1$. For the sake of notation,
we introduce the abbreviations, $m_i=m_{\tilde{\chi}^0_i}$.
The seven quartic charges $N_{1,2,3,4}$ and $N'_{1,2,3}$ for the 3--body
neutralino decays are defined in terms of the bilinear charges by
\begin{eqnarray}
&& N_1=\frac{1}{4}
            \left[|D_{RR}|^2+|D_{LL}|^2
                 +|D_{RL}|^2+|D_{LR}|^2\right]  \nonumber\\
&& N_2=\frac{1}{2}\,\real
            \left(D_{RR}D^{*}_{LR}
                 +D_{LL}D^{*}_{RL}\right)       \nonumber\\
&& N_3=\frac{1}{4}
            \left[|D_{LL}|^2+|D_{RR}|^2
                 -|D_{RL}|^2-|D_{LR}|^2\right]  \nonumber\\
&& N_4=\frac{1}{2}\,\imag
            \left(D_{RR}D^{*}_{LR}
                 +D_{LL}D^{*}_{RL}\right)       \nonumber\\
&& N'_1=\frac{1}{4}
             \left[|D_{RR}|^2+|D_{RL}|^2
                  -|D_{LR}|^2-|D_{LL}|^2\right] \nonumber\\
&& N'_2=\frac{1}{2}\,\real
             \left(D_{RR}D^{*}_{LR}
                  -D_{LL}D^{*}_{RL}\right)      \nonumber\\
&& N'_3=\frac{1}{4}
             \left[|D_{RR}|^2+|D_{LR}|^2
                  -|D_{RL}|^2-|D_{LL}|^2\right]
\end{eqnarray}
The quartic charges $N_{1,2,3,4}$ are P--even,  while the quartic charges
$N'_{1,2,3}$ are P--odd.\s

\begin{figure}[!htb]
\begin{center}
\begin{picture}(350,180)(0,35)
\Text(295,100)[]{$z$}
\DashLine(150,100)(280,100){2}
\Text(150,210)[]{$x$}
\DashLine(150,100)(150,200){2}
\Text(245,190)[]{\color{blue}$\ell^-,\, {\color{red} x_-}$}
\LongArrow(150,100)(200,150)
\DashLine(200,150)(230,180){2}
\Text(90,155)[]{\color{blue}$\ell^+,\, {\color{red} x_+}$}
\LongArrow(150,100)(98,139)
\Text(162,35)[]{\color{blue}$\tilde{\chi}^0_1$}
\LongArrow(150,100)(160,50)
\Text(211,120)[]{\color{red} $\theta$}
\LongArrowArc(150,100)(50,0,45)
\Text(165,145)[]{$\chi$}
\LongArrowArc(150,100)(35,44,145)
\Text(245,90)[]{\color{red} $\hat{n}$}
\LongArrow(150,100)(240,100)
\ArrowArcn(210,160)(15,110,350)
\Text(235,160)[]{\color{red} $\alpha$}
\end{picture}
\end{center}
\caption{\it A schematic diagram of the kinematic configuration of the
             momenta and spin vector in the initial neutralino rest
             mass frame for the three--body neutralino decay
             $\tilde{\chi}^0_i\to \tilde{\chi}^0_1 \ell^+ \ell^-$ in terms
             of two dimensionless energy variables $x_{\pm}$ and two
             angles $\theta$ and $\alpha$. The opening angle $\chi$ is uniquely
             determined by two variables $x_\pm$.  }
\label{fig:kinematical}
\end{figure}
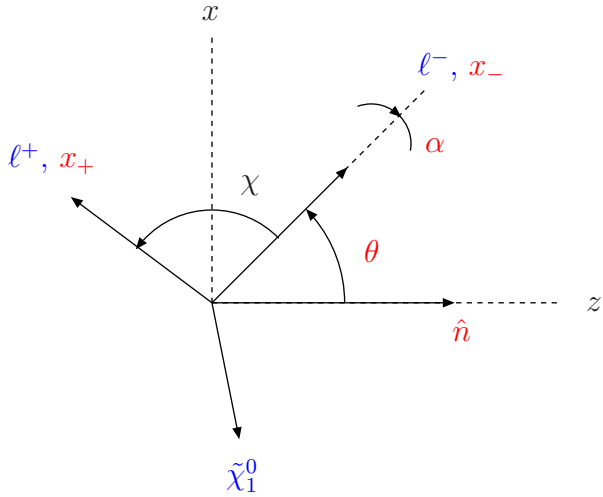

We choose the rest frame of the decaying neutralino as a reference frame to
describe the 3-momenta of the decay products. The neutralino spin 3-vector
$\hat n=(0,0,1)$ defines the direction of the $z$-axis. Since the azimuthal
angle is irrelevant, the 3-momentum vector of the negative lepton can be taken
to
fix the $x$-$z$ plane and its polar angle is denoted by $\theta$ ($0\leq
\theta\leq \pi$).  The orientation of the neutralino decay plane (NDP) is then
fully determined by specifying an additional angle $\alpha$ ($0\leq \alpha\leq
2\pi$), so that by rotating the NDP by $-\alpha$ around the $\ell^-$ direction
it is brought to $x$-$z$ plane, as depicted in Fig.~\ref{fig:kinematical}.  The
differential decay distribution is written in terms of two dimensionless
energy variables, $x_- = 2 E_{e^-}/m_i$ and $x_+ = 2 E_{e^+}/m_i$, and the 
two angles, $\theta$ and $\alpha$, as
\begin{eqnarray}
&& \frac{d^4\Gamma}{dx_- dx_+ d\cos\theta\, d\alpha}
       =\frac{\alpha^2\, m_i}{16 \pi^2}\,\bigg[ F_0(x_-, x_+)
            + (\hat{q}_- \cdot \hat{n})\, F_1(x_-, x_+)
            + (\hat{q}_+ \cdot \hat{n})\, F_2(x_-, x_+)\nonumber\\
       && { }\hskip 4.3cm
            + \hat{n}\cdot(\hat{q}_- \times \hat{q}_+)\,\, F_3(x_-, x_+)\,
            \bigg]
\label{eq:decay rate}
\end{eqnarray}
where $\cos\theta=\hat{q}_-\cdot\hat{n}$, $\hat
q_\pm=\vec{q}_\pm/|\vec{q}_\pm|$ and lepton masses are neglected.  The  
four
kinematic functions $F_k(x_-,x_+)$ ($k = 0$--$3$) are expressed in
terms of the dimensionless energy variables, $x_-$ and $x_+$, and the
quartic charges as
\begin{eqnarray}
&& F_0(x_-,x_+)\,=\, x_- y_-\, (N_1-N_3) + x_+ y_+ \,(N_1+N_3)
                  -2r_{i1} (x_- + x_+ - 1 + r_{i1}^2)\, N_2  \nonumber \\
&& F_1(x_-,x_+)\,=\, -x_- y_-\, (N_1^\prime - N_3^\prime)
                  -r_{i1} x_- x_+\, N_2^\prime            \nonumber  \\
&& F_2(x_-,x_+)\,=\, -x_+ y_+\, (N_1^\prime + N_3^\prime)
                  +r_{i1} x_- x_+\, N_2^\prime           \nonumber  \\
&& F_3(x_-,x_+)\,=\, r_{i1}\, x_- x_+\, N_4
\label{eq:kinematical_functions}
\end{eqnarray}
where $y_\pm=1-x_\pm-r^2_{i1}$ with $r_{i1}=m_1/m_i$.
All  quartic charges
are functions only of the energy variables $x_{\pm}$, but independent of the
orientation angles, $\theta$ and $\alpha$.\s

In the kinematic configuration of Fig.~\ref{fig:kinematical} with
$\hat{n}=(0,0,1)$ taken along the positive $z$--axis and the negative lepton
momentum on the $x$--$z$ plane, the spin dependent scalar
products are expressed in terms of the opening angle, $\chi$, and two angles,
$\theta$ and $\alpha$, as
\begin{eqnarray}
&&\hat{q}_-\cdot\hat{n} = \cos{\theta},\qquad\qquad
\hat{q}_+\cdot\hat{n} = \cos{\chi}\cos{\theta}
                           -\sin{\chi}\sin{\theta}\cos{\alpha}\nonumber \\
&&\hat{n}\cdot(\hat{q}_-\times\hat{q}_+) = \sin{\chi}\sin{\theta}\sin{\alpha}
\end{eqnarray}
where the cosine of two lepton momentum directions, $\cos{\chi}$, is simply a
function of two normalized energy variables, $x_-$ and $x_+$:
\begin{eqnarray}
\cos{\chi} =1-2\,\, (x_- + x_+ - 1+ r_{i1}^2)/(x_- x_+)
\end{eqnarray}
and $\sin\chi = \sqrt{1-\cos^2\chi}$.
The kinematically--allowed ranges for the angles, $\theta$ and $\alpha$, and
the dimensionless energy variables, $x_+$ and $x_-$, are
\begin{eqnarray}
0 \leq \theta,\, \frac{\alpha}{2} \leq\pi, \ \
0 \leq x_\pm  \leq 1-r_{i1}^2,\ \
  (1-x_-)(1-x_+)\geq r_{i1}^2,\ \ x_- + x_+ \geq 1-r_{i1}^2
\label{eq:kinematical_range}
\end{eqnarray}
The allowed crescent--shaped $x_+$ and $x_-$ region (see the left panel of
Fig.~\ref{fig:dalitz}) is restricted due to
four--momentum conservation, while the full ranges of the angles $\theta$ and
$\alpha$ allowed.\s

\subsection{Majorana nature and implications of CP transformation}

Before presenting a few concrete numerical examples for probing the Majorana
nature and CP violation in the neutralino system through the leptonic
  three--body
neutralino decays, we investigate the
implications of the invariance under CP and CP$\tilde{\rm T}$
transformations\footnote{The naive time reversal transformation $\tilde{\rm T}$
reverses the direction of all 3--momenta and spins, but does
not exchange initial and final
states. Quantities that are odd under CP$\tilde{\rm T}$ can be non--zero only
for complex transition amplitudes with absorptive phases which can be
generated,
for example, by loops, and Breit--Wigner propagators.} for the three
body leptonic neutralino decays \cite{Bilenky,Moortgat-Pick1}.\s

In the rest frame of the decaying neutralino, three final
particles form a decay plane and every three--momentum changes its sign
under P transformation as well as $\tilde{\rm T}$ transformation while
C transformation exchanges the four momenta of two leptons. The polarization
vector $\hat{n}$ does not change under P and C transformations, but it changes
its
sign under $\tilde{\rm T}$ transformation. Then, the CP operation
transforms the momenta and spin vector in the decay processes as
\begin{eqnarray}
x_\pm \rightarrow +x_\mp,\ \
   \vec{q}_\pm \rightarrow -\vec{q}_\mp, \ \ \hat{q}_\pm\cdot \hat{n}
   \rightarrow -\hat{q}_\mp\cdot \hat{n},\ \
  \hat{n}\cdot(\hat{q}_-\times\hat{q}_+) \rightarrow
  -\hat{n}\cdot(\hat{q}_-\times\hat{q}_+)
\end{eqnarray}
while the CP$\tilde{\rm T}$ operations transform the momentum and spin vectors
as
\begin{eqnarray}
x_\pm \rightarrow +x_\mp,\ \ \vec{q}_\pm \rightarrow +\vec{q}_\mp,\ \
   \hat{q}_\pm\cdot \hat{n} \rightarrow - \hat{q}_\mp\cdot \hat{n}, \ \
   \hat{n}\cdot(\hat{q}_-\times\hat{q}_+) \rightarrow
   +\hat{n}\cdot(\hat{q}_-\times\hat{q}_+)
\end{eqnarray}
with the condition that the complex conjugation of the decay amplitude is
taken \cite{Naive_T}.\s

Since neutralinos are  the Majorana particles,
CP invariance in the three--body neutralino decay
$\tilde{\chi}^0_i\to \tilde{\chi}^0_j \ell^+\ell^-$ leads to
additional relations
among the
bilinear charges defined in Eq.(\ref{eq:bilinear_charges}):
\begin{eqnarray}
&& D_{LR} = \eta_i \eta_j D_{RR}(t\leftrightarrow u)\nonumber\\
&& D_{RL} = \eta_i \eta_j D_{LL}(t\leftrightarrow u)
\end{eqnarray}
where $\eta_{i,j}=\pm i$ are the intrinsic CP parities \cite{Kayser} of the
neutralinos, $\tilde{\chi}^0_{i,j}$, respectively, and, as a result,
to the CP relations for the kinematic functions defined in
Eq.(\ref{eq:kinematical_functions}):
\begin{eqnarray}
&& F_0(x_-,x_+) = +F_0(x_+,x_-)\nonumber\\
&& F_1(x_-,x_+) = -F_2(x_+,x_-)\nonumber\\
&& F_3(x_-,x_+) = -F_3(x_+,x_-)
\label{eq:cp_relation}
\end{eqnarray}
in the three--body leptonic neutralino decays. We note that the CP relations
(\ref{eq:cp_relation}) can be satisfied only when the couplings ${\cal
  Z}_{ij}$, $g_{Lij}$ and $g_{Rij}$ are {\it simultaneously} purely real (for
$\eta_i=\eta_j =\pm i$) or {\it simultaneously} purely imaginary (for
$\eta_i=-\eta_j=\pm i$).\s

On the other hand, CP$\tilde{\rm T}$ invariance leads to the relations among
the bilinear charges:
\begin{eqnarray}
&& D_{LR} = - D^*_{RR}(t\leftrightarrow u)\nonumber\\
&& D_{RL} = - D^*_{LL}(t\leftrightarrow u)
\end{eqnarray}
These CP$\tilde{\rm T}$ relations are satisfied if the $Z$--boson and slepton
widths
are neglected, that is to say, if there are no absorptive parts in the process.
In the approximation of neglecting particle widths, we have the following
CP$\tilde{\rm T}$ relations
for the kinematic functions:
\begin{eqnarray}
&& F_0(x_-,x_+) = +F_0(x_+,x_-)\nonumber\\
&& F_1(x_-,x_+) = -F_2(x_+,x_-)\nonumber\\
&& F_3(x_-,x_+) = +F_3(x_+,x_-)
\label{eq:cpt_relation}
\end{eqnarray}
{\it independently} of the mixing character of neutralinos and whether CP
is violated or not.  

CP--conserving absorptive parts appear in loop diagrams with on--shell
propagators through final--state interactions or from the widths of
intermediate unstable particles. In addition, CP$\tilde{\rm T}$--odd
asymmetries may arise from the interference between a dominant tree--level and
a sub--leading loop diagram mediating the decay. However, those absorptive
parts and interference effects are usually tiny in the leptonic decay
involving only electroweak interactions, at most at a level of a couple
  of percents. In this light, we will ignore all the
width effects and electroweak loop corrections in the following analytic and
numerical analyses.\s

\section{Numerical Analyses}
\label{sec:majorana_cp}

In the numerical analyses below we adopt an mSUGRA scenario defined by
\begin{eqnarray}
&& m_0=150\; {\rm GeV}, \quad m_{1/2}=200\; {\rm  GeV}, \quad
 A_0=-650\; {\rm GeV}
\end{eqnarray}
at the GUT scale requiring the pole mass of the top quark $ m_t=178$ GeV, and
\begin{eqnarray}
&& \tan\beta=10, \qquad\quad
{\rm sgn}(\mu)>0
\end{eqnarray}
at the electroweak scale at which all  parameters are derived
with the RGE code SPheno \cite{SPheno} (very similar results are obtained with
other RGE codes; for comparison of different codes see \cite{Kraml}).
For the light neutralino and chargino masses we find
\begin{eqnarray}
m_{\tilde{\chi}^0_1}=78.1\,{\rm GeV}, \qquad
m_{\tilde{\chi}^0_2}=148.5\,{\rm GeV}, \qquad
m_{\tilde{\chi}^\pm_1}=148.4\,{\rm GeV}
\label{eq:light chimasses}
\end{eqnarray}
and we note that the light neutralino and chargino masses and mixing angles
are reproduced fairly well with the tree-level formulae taking
\begin{eqnarray}
M_1=80\, {\rm GeV},\ \ M_2=158\, {\rm GeV},\ \ |\mu|=415\, {\rm GeV},\ \
\Phi_\mu=0; \ \ \tan\beta=10
\label{eq:parameter_set}
\end{eqnarray}
The selectron and sneutrino masses derived with the RGE code SPheno are
\begin{eqnarray}
   m_{\tilde{e}_{_L}}=207.7\,{\rm GeV},    \qquad
   m_{\tilde{e}_{_R}}=173.1\,{\rm GeV},   \qquad
   m_{\tilde{\nu}_e}=192.1\,{\rm GeV}
\label{eq:semasses}
\end{eqnarray}
Relevant for our analysis, the derived branching ratios for leptonic
three--body
decays of the second lightest neutralino are
\begin{eqnarray}
{\rm Br}(\tilde\chi^0_2\to \tilde\chi^0_1e^+e^-)=4.5\%, \qquad
{\rm Br}(\tilde\chi^0_2\to \tilde\chi^0_1\mu^+\mu^-)=4.6\%
\label{chiBR}
\end{eqnarray}
and for the $\tilde{e}_L\to  \tilde\chi^0_2 e$ decay
\begin{eqnarray}
{\rm Br}(\tilde{e}_L\to  \tilde\chi^0_2e)=28.4\%
\label{selBR}
\end{eqnarray}
while other decay modes have
${\rm Br}(\tilde\chi^0_2\to \tilde\chi^0_1\tau^+\tau^-)=58.9\%$,
${\rm Br}(\tilde\chi^0_2\to \tilde\chi^0_1 q\bar{q} )=9.1\%$,
${\rm Br}(\tilde\chi^0_2\to \tilde\chi^0_1 \nu\bar{\nu} )=23.7\%$,
${\rm Br}(\tilde{e}_L\to  \tilde\chi^0_1e)=21.4\%$,
${\rm Br}(\tilde{e}_L\to  \tilde\chi^-_1\nu)=50.3\%$,
${\rm Br}(\tilde{\nu}_e\to\tilde\chi^0_2\nu)=19.6\%$,
${\rm Br}(\tilde{\nu}_e\to\tilde\chi^+_1 e)=45\%$,
and
${\rm Br}(\tilde{t}_1\to \bar{b} \tilde\chi^+_1)=98.3\%$,
${\rm Br}(\tilde{t}_1\to c \tilde\chi^0_2)=1.6\%$.\s

The production cross sections for selectron-pair production
processes with unpolarized $e^+e^-$ beams at $\sqrt{s}=500$ GeV
are as follows
\begin{eqnarray}
&& \sigma\{\tilde{e}^+_R\tilde{e}^-_R\} = 273.4\, {\rm fb}, \quad
   \sigma\{\tilde{e}^\pm_R\tilde{e}^\mp_L\} = 113.5\, {\rm fb}, \quad
   \sigma\{\tilde{e}^+_L\tilde{e}^-_L\} = 80.7\, {\rm fb}
\end{eqnarray}
so that large ensembles of events,  $\sim 2\times 10^5 $ events
for $ \tilde{e}^\pm_R\tilde{e}^\mp_L$ and
$ \tilde{e}^+_L\tilde{e}^-_L$ at integrated luminosity of 1000
fb$^{-1}$,  will be generated. Given the branching fractions in
Eqs.~(\ref{chiBR}) and (\ref{selBR}), a sufficient number of
events for the decays  $\tilde{\chi}^0_{2}\to\tilde{\chi}^0_{1} e^+e^-$ and
$\tilde{\chi}^0_2\to\tilde{\chi}^0_{1} \mu^+\mu^-$ are expected to be
selected\footnote{In Monte Carlo simulations we conservatively
assume that at least 1000 neutralino decay  events can be selected
and we evaluate all the relevant physical quantities with their tree--level
formulas based on the parameter set (\ref{eq:parameter_set}).}, allowing
the analysis of the properties of the neutralino decay at great detail.\s

For completeness, we note that other production channels for SUSY-related
processes that are
either open or with cross sections greater than 1 fb at $\sqrt{s}=500$ GeV
have the following cross sections:
$\sigma\{\tilde{\chi}^0_1\tilde{\chi}^0_1\} = 264.8$  fb, 
$\sigma\{\tilde{\chi}^0_1\tilde{\chi}^0_2\} = 159.1$ fb, 
$\sigma\{\tilde{\chi}^0_2\tilde{\chi}^0_2\} = 116.8$ fb,  
$\sigma\{\tilde{\chi}^+_1\tilde{\chi}^-_1\} = 294$ fb,
$\sigma\{\tilde{\mu}^+_R\tilde{\mu}^-_R\} = 39.8$ fb,
$\sigma\{\tilde{\mu}^+_L\tilde{\mu}^-_L\} = 22.3$ fb,
$\sigma\{\tilde{\tau}^+_1\tilde{\tau}^-_1\} = 52.4$ fb,
$\sigma\{\tilde{\tau}^\pm_1\tilde{\tau}^\mp_2\} = 5.2$ fb,
$\sigma\{\tilde{\tau}^+_2\tilde{\tau}^-_2\} = 17.1$ fb,
$\sigma\{\tilde{\nu}_e\tilde{\nu}^*_e\}=662.2$ fb,
$\sigma\{\tilde{\nu}_\mu\tilde{\nu}^*_\mu\}=15.2$ fb,
$\sigma\{\tilde{\nu}_\tau\tilde{\nu}^*_\tau\}=16.8$ fb,
$\sigma\{\tilde{t}_1\tilde{t}^*_1\} = 45$ fb and
$\sigma\{ h^0 Z\} = 64.6$ fb.
Most of these processes can be separated by simple
kinematical cuts. Note that the $\tilde{\nu}_e\tilde{\nu}^*_e$
production
process  might be exploited for our purposes if the decay mode $\tilde\chi^+_1
e$ of
$\tilde{\nu}_e$ in one hemisphere could be used to tag the decay mode
$\tilde{\chi}^0_2\nu$ of the second electron--sneutrino in the other
hemisphere. However, since 
the detailed experimental simulations to assess this possibility are beyond the
scope of the paper, we do not include the $\tilde{\nu}_e\tilde{\nu}^*_e$
channel in our analyses.

\subsection{Lepton energy distribution}

The polarization--independent kinematic function $F_0$ is symmetric with
respect to the energy variables $x_+$ and $x_-$ exactly in the CP invariant
case and to a good approximation in the CP non--invariant case. 
One
of the most decisive ways for confirming the Majorana nature of the
neutralinos, namely  
the observation of the symmetric distribution of events on the
$(x_-, x_+)$ Dalitz plane \cite{Bilenky}, can therefore be realized  
in the cascade decay process  of Eq.~(\ref{cascade}) in which the 
$\tilde{\chi}^0_2$ rest frame is reconstructable.\s

\begin{figure}[ht!]
\begin{center}
\includegraphics[height=7.5cm,width=7.5cm]{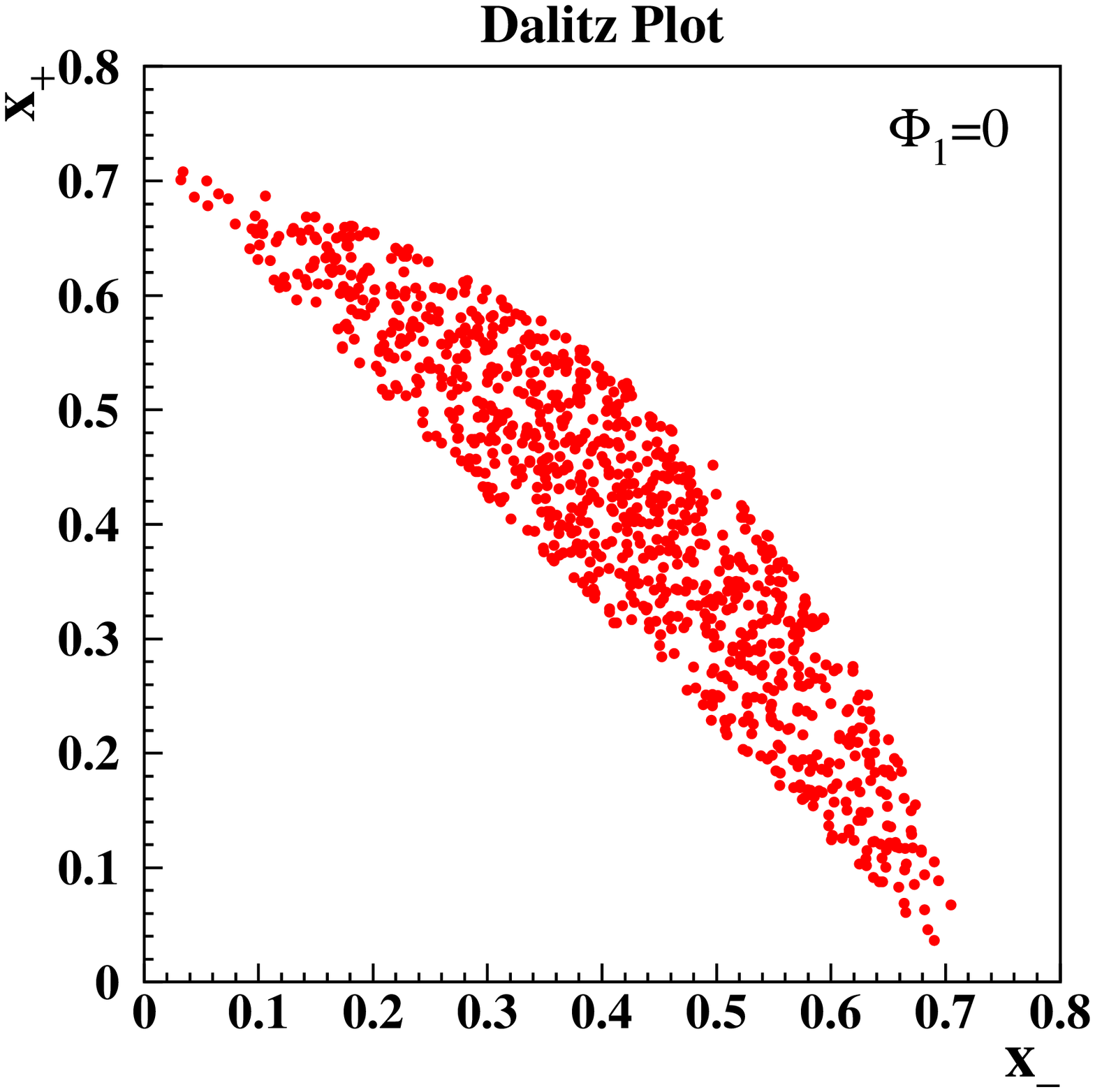}\hskip 0.5cm
\includegraphics[height=7.5cm,width=7.5cm]{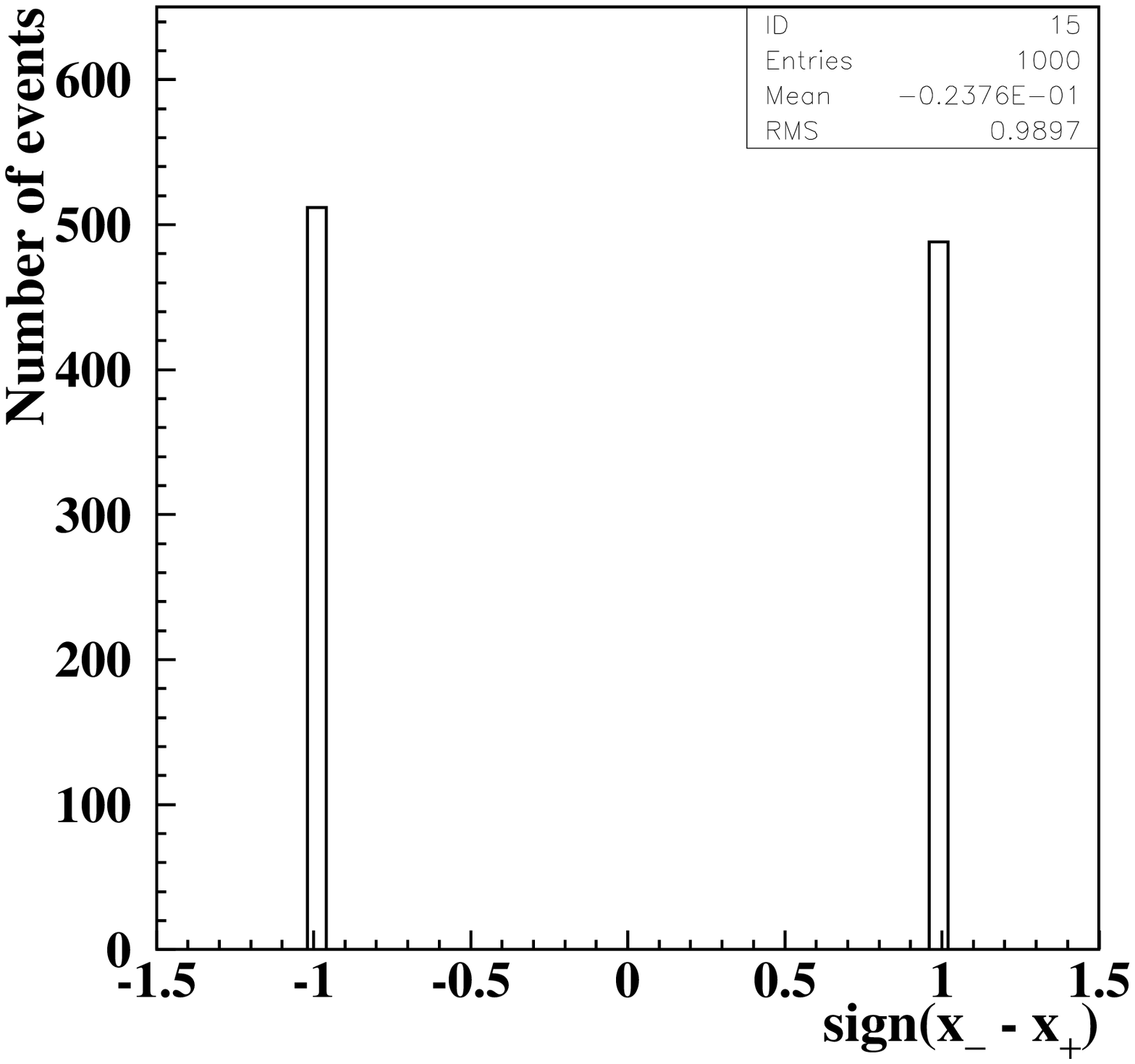}
\end{center}
\vskip -0.5cm
\caption{\it Left: the Dalitz plot of the neutralino
  decay $\tilde{\chi}^0_2\to\tilde{\chi}^0_1\ell^+\ell^-$ in the $(x_-,x_+)$
  Dalitz plane. Right: the number of events with ${\rm sign}(x_--x_+) = - $
  (left histogram) and ${\rm sign}(x_--x_+)=+$ (right histogram).  $\Phi_1=0$
  is taken and the parameter set (\ref{eq:parameter_set}) is used for the
  other relevant SUSY parameters. The event distribution is generated with the
  total number of events of 1000 by a Monte Carlo method. }
\label{fig:dalitz}
\end{figure}

The left panel of Fig.~\ref{fig:dalitz} shows the Dalitz plot of the leptonic
three--body decay $\tilde{\chi}^0_2\to\tilde{\chi}^0_1 \ell^+\ell^-$ in the
$(x_-,x_+)$ plane for the parameter set (\ref{eq:parameter_set}) with
$\Phi_1=0$. The right panel shows the numbers of piled--up events with
${\rm sign}(x_--x_+)=-$ (left histogram) and ${\rm sign}(x_--x_+)=+$ (right
histogram), simulated with 1000 events by a Monte Carlo method. We
note that the difference of two numbers of events $\Delta N_{\rm ev}\approx
24$ is within the expected statistical error of $\Delta N_{\rm
exp}=\sqrt{N_{\rm ev}}\simeq 32$ with $N_{\rm ev}=1000$. \s

\subsection{Lepton angular distribution}

Another kinematic distribution of great interest 
is the lepton angle distribution with respect to the neutralino
polarization 
vector; the lepton angle distribution with respect to the beam direction 
in the $e^+e^-$ reference frame has been discussed in \cite{Moortgat-Pick1}.
Defining $\theta_\pm$ to be the polar angle between the $\ell^\pm$
momentum and polarization vector $\hat{n}$, the normalized lepton angle
distribution can be written as
\begin{eqnarray}
\frac{1}{\Gamma}\frac{d\Gamma}{d\cos\theta_\pm}
  = \frac{1}{2} \left(1\, \pm\, \xi_\pm\, \cos\theta_\pm \right)
\label{eq:lepton_angle}
\end{eqnarray}
with $\cos\theta_\pm\equiv\hat{q}_\pm\cdot \hat{n}$. In the
approximation of all particle widths neglected,  the slope parameters have
  to be equal, $\xi_-=\xi_+$, {\it irrespective of whether the theory is
CP invariant or not}. This is the consequence  
of CP$\tilde{\rm T}$ invariance  of the decay distribution
of the Majorana particle, 
cf. the second relation of Eq.~(\ref{eq:cpt_relation}).  
As a result, the
sum of two lepton angle distributions has to be 
flat, that is to say, independent
of the polar angles. This is one of the genuine tests of the Majorana
nature of the neutralinos.  \s

\begin{figure}[ht!]
\begin{center}
\epsfig{figure=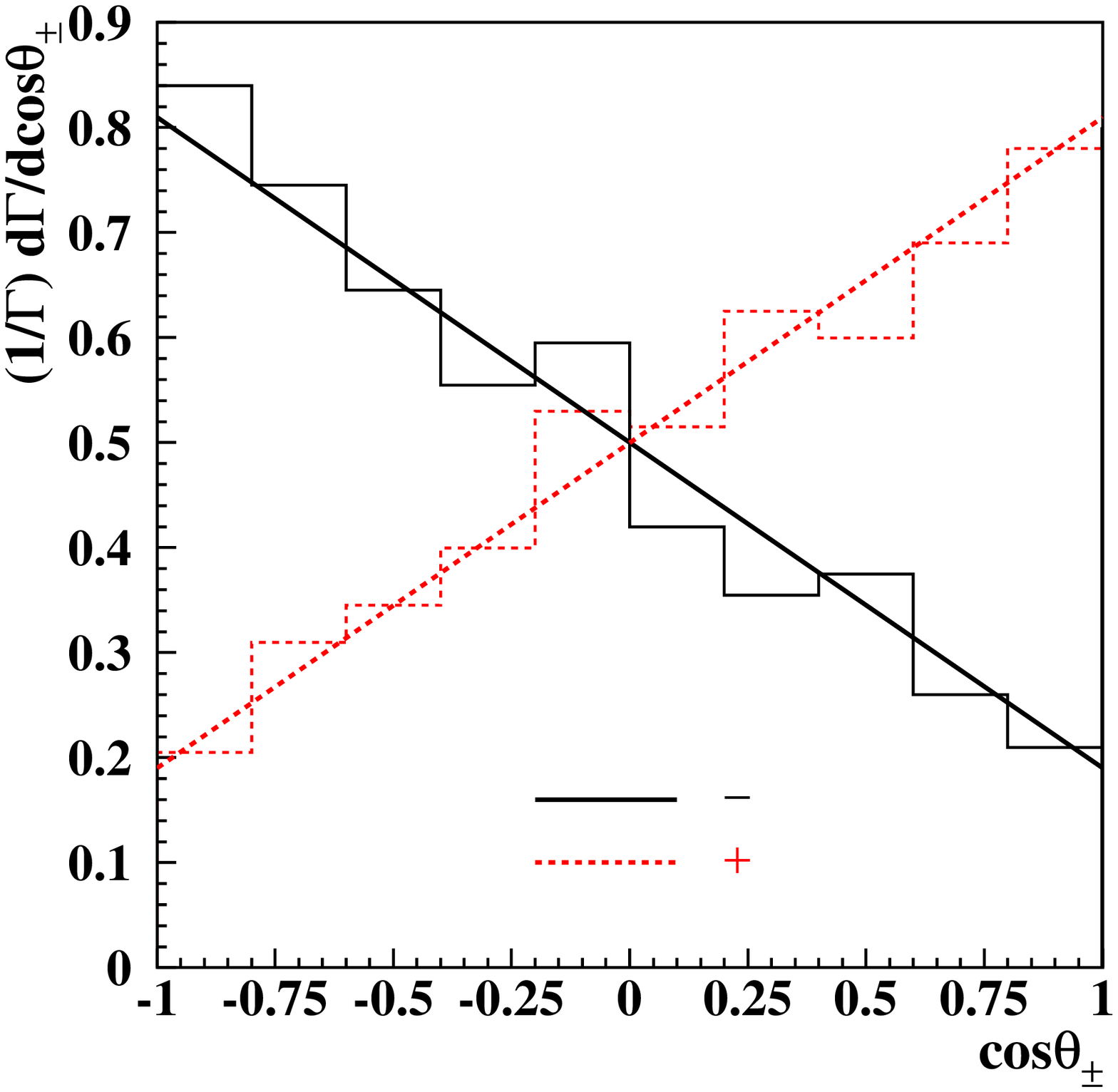,width=7.5cm,height=7.5cm}\hskip 0.5cm
\epsfig{figure=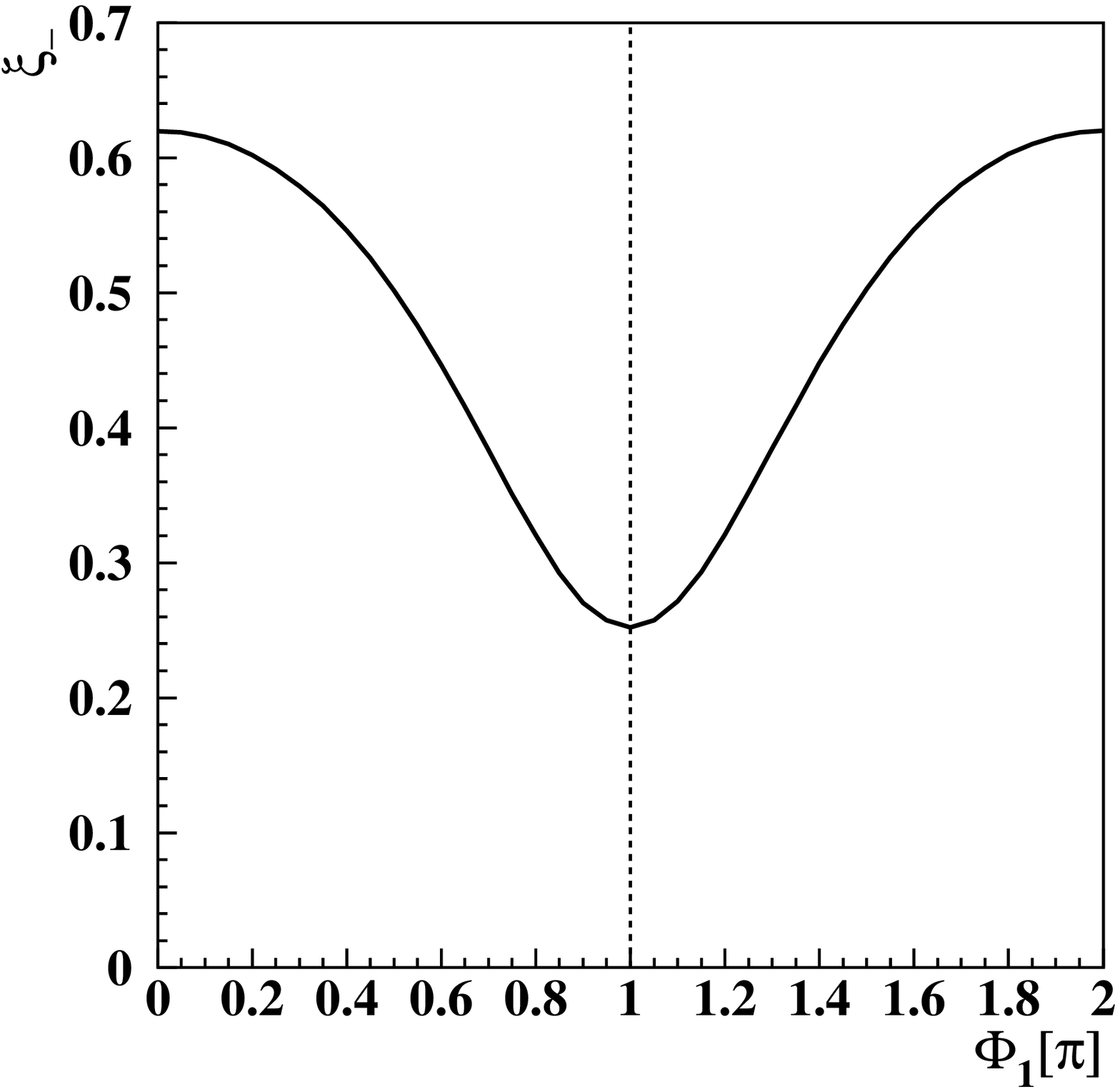,width=7.5cm,height=7.5cm}
\end{center}
\vskip -0.5cm
\caption{\it Left: The normalized lepton angle distribution
  (\ref{eq:lepton_angle})
  of the neutralino decay $\tilde{\chi}^0_2\to\tilde{\chi}^0_1\ell^+\ell^-$;
  the black solid line is for the negative charged lepton and the red
  dashed line for
  the positive charged lepton. Right: The $\Phi_1$ dependence of the slope
  parameter $\xi_-$ for the parameter set (\ref{eq:parameter_set}).}
\label{fig:etas}
\end{figure}

The left panel of Fig.~\ref{fig:etas} shows the lepton angle
distribution for the parameter set (\ref{eq:parameter_set}) with the phase
$\Phi_1=0$.  The solid line is for the cosine of the negatively--charged
lepton angle, $\cos\theta_-$, and the dashed line for the cosine of the
positively--charged lepton angle, $\cos\theta_+$. 
A simple numerical analysis based on the number
of events of $N_{\rm ev}=1000$ shows that the CP$\tilde{\rm T}$ relation
can be confirmed  within 1--$\sigma$ statistical uncertainty of about 8\%
for the whole range of $\cos\theta_\pm$. It increases to
about 10\%  for the range
with $|\cos\theta_\pm |> 0.8$ vetoed if the cut turns out
necessary to avoid distortions of  the $\ell^\pm$ distributions
differently by experimental selection criteria \cite{Aguilar3}.
Certainly, it will be
necessary to perform a more detailed and realistic experimental analysis
including all the systematic uncertainties before drawing a more concrete
conclusion. Such a  comprehensive analysis is, however, beyond the scope
of the present theoretical investigations.\s

The quantities $\xi_\pm$, denoting the slope of the lepton angle distribution
with respect to $\cos\theta_\pm$, depend on the values of the SUSY parameters.
In the right panel of
Fig.~\ref{fig:etas} the $\Phi_1$ dependence of the CP--even
quantities $\xi_\pm$ is shown demonstrating that the $\xi_\pm$ measurement
can provide information on the CP-violating phase.
\s

\subsection{Lepton invariant mass and opening--angle distribution}

The invariant mass $m_{ll}$ of two final--state leptons in the decay
$\tilde{\chi}^0_2\to\tilde{\chi}^0_1\ell^+\ell^-$ is a Lorentz--invariant
kinematic variable so that it is straightforward to reconstruct the quantity
experimentally by measuring the four momenta of two final--state leptons in
the laboratory frame. Furthermore, the distribution for the invariant mass
$m_{ll}$ is independent of the specific production process for the decaying
neutralino, because the invariant mass distribution does not involve any spin
correlations between production and decay \cite{Moortgat-Pick1}.\s

Near the maximum
end point of the lepton invariant mass distribution, the neutralino
$\tilde{\chi}^0_1$ is produced nearly at rest, the Mandelstam variable
$s$ approaches to $(m_2-m_1)^2$ and the variables $t$ and $u$ become identical
$\sim m_1 m_2$. Ignoring the particle widths in the propagators we find from
Eq.~(\ref{eq:neutralino decay amplitude}) that the decay amplitude can be
written approximately as
\begin{eqnarray}
{\cal D} \, \sim\, \frac{1}{m_2}\bar{u}(\tilde{\chi}^0_1)\,
            \bigg\{\!\!\not\!{L}\left[\, i \imag(X_L)
            +\gamma_5\real(X_L)\right]+ \not\!{R} \left[ i \imag(X_R)
            +\gamma_5\real(X_R)\right]\bigg\}\, u(\tilde{\chi}^0_2)
\end{eqnarray}
where $X_L=D_{LL}$ and $X_R=D_{LR}$ near the invariant mass end point
and $L^\mu/R^\mu = \bar{u}(\ell^-)\gamma^\mu P_{L,R}\,\, v(\ell^+)$ are  the
left--handed/right--handed lepton vector currents. The approximate form of
the decay amplitude leads to the absolute amplitude squared of the following
approximate form:
\begin{eqnarray}
|{\cal D}|^2\, \sim\, r_{21} \left(1-r_{21}\right)^2\,
    \left\{\,\real(X_L)^2 + \real(X_R)^2\,\right\} + O(\beta^2)
\end{eqnarray}
near the end point with the neutralino $\tilde{\chi}^0_1$ velocity $\beta$
$\sim \sqrt{1-\mu_{ll}}$ for the normalized invariant mass,
$\mu_{ll}=m_{ll}/(m_2-m_1)$.  Therefore, the invariant mass distribution
exhibits a characteristic steep $S$-wave (slow $P$-wave)
decrease proportional to $\beta$
($\beta^3$) when the CP parities of two neutralinos are the same
(opposite), that is to say, if $X_{L,R}$ is purely real (purely
imaginary). On the other hand, in the CP non-invariant case, where both the real
and imaginary parts of $X_{L,R}$ are non--vanishing,
the invariant mass distribution decreases
steeply in $S$--wave \cite{SYChoi1}.\s

The threshold behavior of the invariant mass distribution can be understood by
investigating the selection rule of the orbital angular momentum by CP
symmetry.  In the non--relativistic limit of two neutralinos, the orbital
angular momentum $L$ of the final two--lepton and LSP system satisfies the CP
relation
\begin{eqnarray}
1=-\eta_1\eta_2 \left(-1\right)^L
\label{eq:selection_rule}
\end{eqnarray}
With the same (opposite) relative CP parity leading to $\eta_1\eta_2=-1$
($+1$), the selection rule (\ref{eq:selection_rule}) forces the final
two--lepton and LSP system to have $L=0$ ($1$) near the threshold. In other
words, the neutralino axial--vector (vector) current corresponds to $S$--wave
($P$--wave) excitations.\s

Additional clear signature of the selection rule (\ref{eq:selection_rule})
is provided by the decay distribution with respect to the opening angle $\chi$
between two leptons \cite{Moortgat-Pick1}.  Since the relation between the
invariant mass and the opening angle is given by
\begin{eqnarray}
m^2_{ll}=\frac{m^2_2}{2}\, x_+ x_- \,\left(1-\cos\chi\right)
\end{eqnarray}
the invariant mass $m_{ll}$ takes its maximum value for $\chi=\pi$ for given
$x_+$ and $x_-$, {\it i.e.} when the momentum directions of two leptons are
opposite.  In this case, the helicities of two leptons coupled to  a
vector current are opposite, rendering its total spin sum to be unity along
the flight direction. Therefore, angular momentum conservation forces the
orbital angular momentum to be zero.  As a consequence, when two neutralinos
have the same (opposite) CP parity, the opening angle distribution is enhanced
(suppressed) near $\cos\chi=-1$.  A similar argument based on total angular
momentum conservation can be applied to show that the opening angle
distribution exhibits the opposite property for $\cos\chi=1$, i.e. for the
vanishing opening angle, $\chi=0$. \s

\begin{figure}[!htb]
\begin{center}
\epsfig{figure=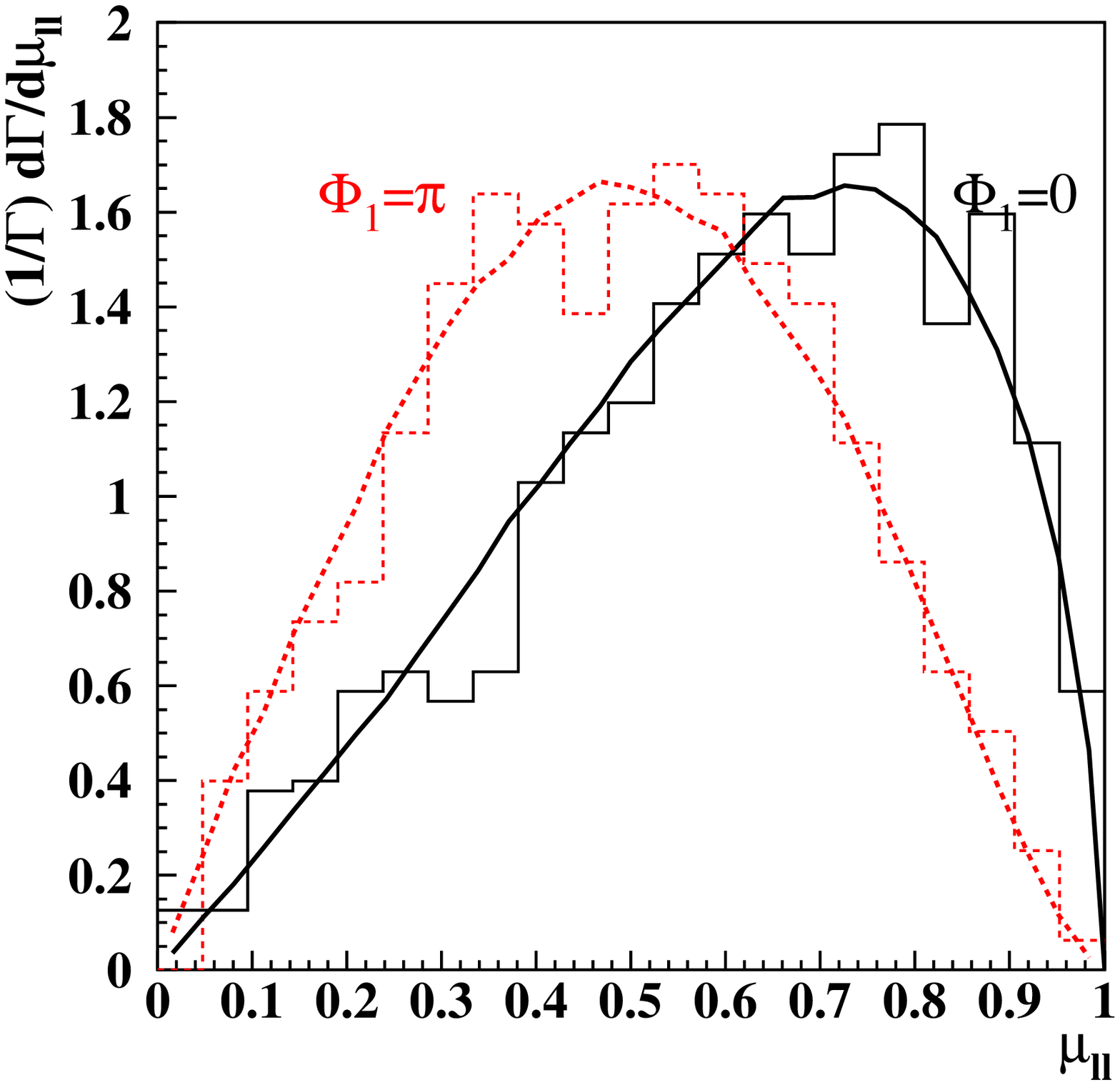,width=7.5cm,height=7.5cm}\hskip 0.5cm
\epsfig{figure=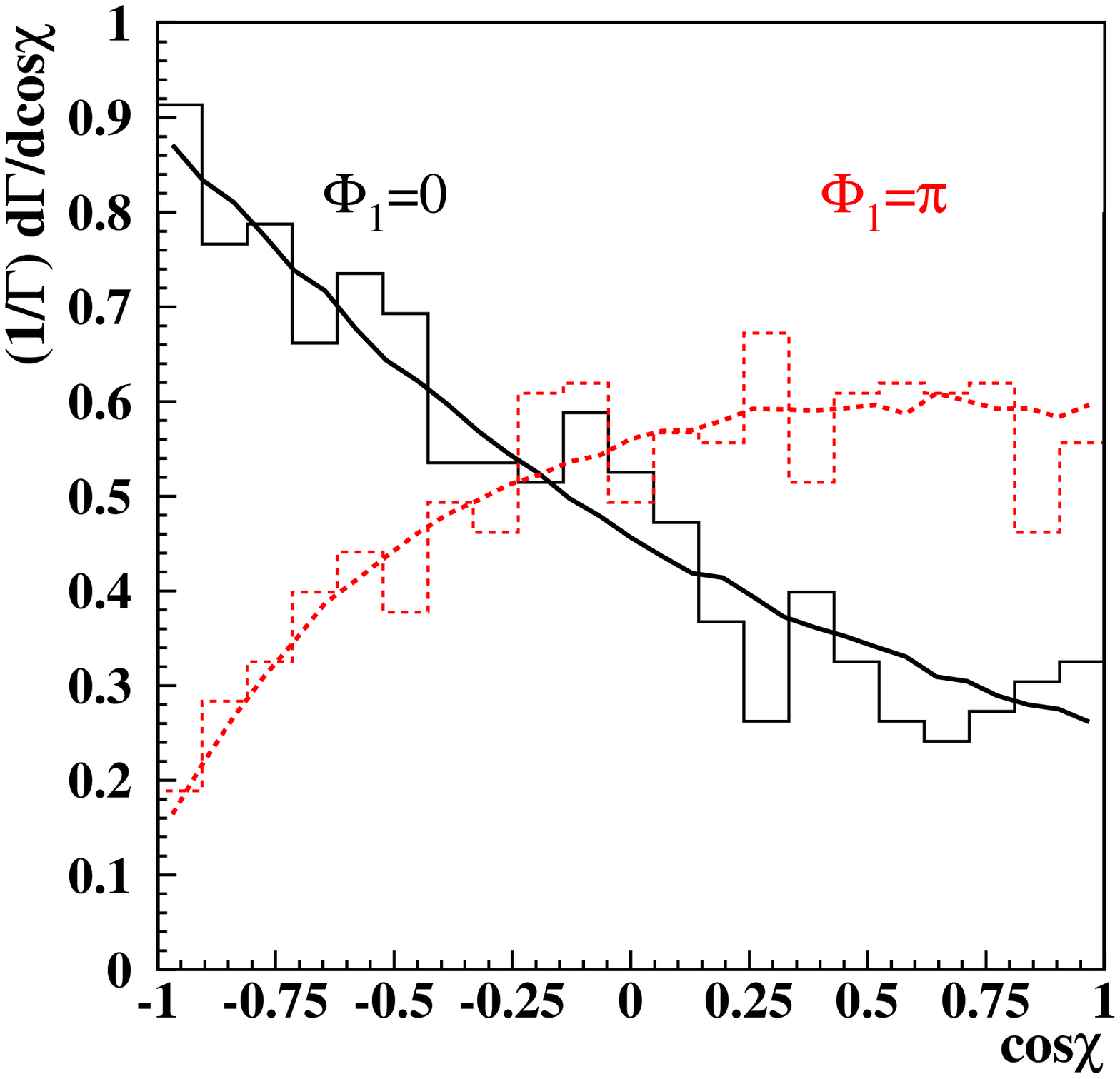,width=7.5cm,height=7.5cm}
\end{center}
\vskip -0.5cm
\caption{\it The lepton normalized invariant mass distribution (left panel) and the
  opening angle distribution (right panel) in the three--body leptonic
  neutralino decay $\tilde{\chi}^0_2\to\tilde{\chi}^0_1\ell^+\ell^-$. The
  black solid lines are for $\Phi_1=0$, (i.e. neutralinos with the same CP
  parities) and the red dashed lines for $\Phi_1=\pi$, (i.e. neutralinos with
  the
  opposite CP parities).}
\label{fig:mass_angle}
\end{figure}

Based on the parameter set (\ref{eq:parameter_set}), we show
in Fig.~\ref{fig:mass_angle} the lepton normalized invariant mass
distribution (left panel) and the opening--angle distribution
(right panel) for $\Phi_1=0$ (black solid line) and $\Phi_1=\pi$
(red dashed line). The histograms are based on 1000 events
generated by Monte Carlo.  The invariant mass distribution
decreases steeply (slowly) near the end point and the opening
angle distribution is strongly enhanced (suppressed) near
$\chi=\pi$ for two neutralinos of the same (opposite) CP parities
in the CP invariant case. Note however, that since in both cases the invariant 
mass distribution vanishes at the end point, the distinction between 
the $S$- and $P$-wave behavior might be tricky. On the other hand, the
opening angle distribution can have a finite value (depending on the relative
CP parity) facilitating the discrimination. In our numerical case,    
the $\chi^2/n$ (with $n=20$ degrees of freedom)
for the fits with $\Phi_1=0/\pi$, respectively, are as follows: 0.86/0.64
for the lepton invariant mass distributions (left panel), and 0.83/0.54 for
the opening angle distributions (right panel), indicating that the theoretical
curves fit the Monte--Carlo generated histograms with very good precision.
Consequently, the invariant mass and/or opening angle distributions can
provide us with a very powerful handle for determining the 
relative CP parity of two
neutralino states $\tilde{\chi}^0_1$ and $\tilde{\chi}^0_2$.\s

\subsection{CP--odd triple spin/momentum product}

The CP and CP$\tilde{\rm T}$ relations in Eqs.~(\ref{eq:cp_relation}) and
(\ref{eq:cpt_relation}) enable us to construct a genuine CP--odd and
CP$\tilde{\rm T}$--even distribution\footnote{In the presence of absorptive
CP--conserving phases due to particle widths or radiative corrections, there
exist 
two additional CP--odd distributions; $F_0(x_-,x_+)-F_0(x_+,x_-)$ and
$F_1(x_-,x_+)+F_2(x_+,x_-)$.}:
\begin{eqnarray}
F_{\rm CP}(x_-,x_+) = \frac{1}{2}\left[\, F_3(x_-,x_+)+F_3(x_+,x_-)\,\right]
\label{eq:cp_asymmetry}
\end{eqnarray}
A typical CP--odd observable related to the CP--odd distribution
(\ref{eq:cp_asymmetry}) is the triple product
\begin{eqnarray}
O_{\rm CP}=\hat{n}\cdot(\hat{q}_+\times\hat{q}_-)
\end{eqnarray}
formed with the spin vector $\hat{n}$ and two final lepton momentum directions.
With the CP--odd observable we can construct the CP--odd asymmetry as follows:
\begin{eqnarray}
A_{\rm CP} \equiv \frac{N(O_{\rm CP} >0) - N(O_{\rm CP} <0)}{
                  N(O_{\rm CP} >0) + N(O_{\rm CP} <0)}
   =\frac{\int_{\cal D} \frac{1}{2} \sin\chi \, F_{\rm CP}(x_-,x_+)\, dx_-
                  dx_+}{
                       \int_{\cal D} F_0(x_-,x_+)\, dx_- dx_+}
\label{eq:cp_odd_asymmetry}
\end{eqnarray}
where ${\cal D}$ denotes the kinematically allowed $(x_-,x_+)$ Dalitz region
defined in Eq.~(\ref{eq:kinematical_range}).\s

\begin{figure}[t]
\begin{center}
\epsfig{figure=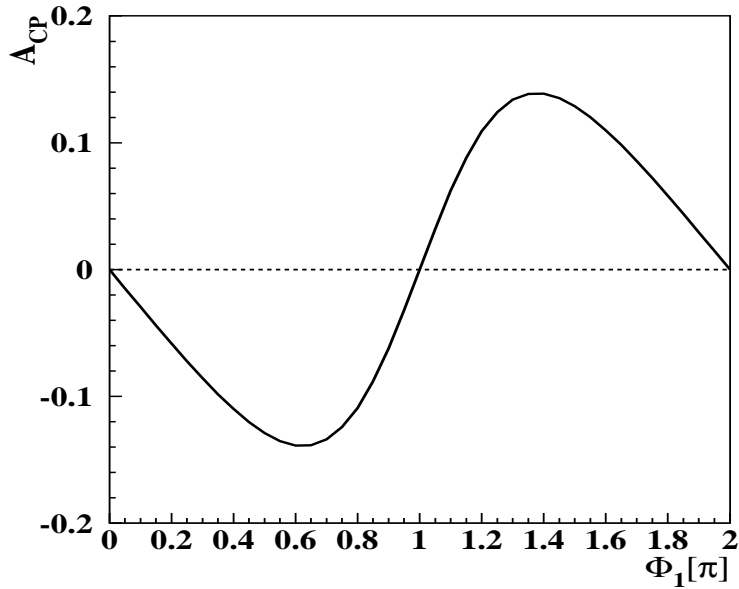,width=11cm,height=9.cm}
\end{center}
\vskip -0.5cm
\caption{\it The $\Phi_1$ dependence of the CP--odd and CP$\tilde{T}$--even
  asymmetry $A_{\rm CP}$ of the triple scalar product for the parameter set
  (\ref{eq:parameter_set}). We note that the 1--$\sigma$ statistical
  uncertainty of the CP--odd asymmetry is
  $\sqrt{(1-A^2_{\rm CP})/N_{\rm ev}}\simeq 3.1\% $ with the number of
  events of $N_{\rm ev}= 1000$. }
\label{fig:triple_product}
\end{figure}

As described in Sect.~\ref{sec:mixing}, CP violation in the neutralino system
arises when the phases of $M_1$ and/or $\mu$ are different from 0 and/or
$\pi$.  These phases generally lead to large contributions to the EDMs.
For the selectron masses under consideration in the present
work, the experimental bounds on the electron EDM put the strongest
constraints on the phases $\Phi_1$ and $\Phi_\mu$. Nevertheless, it has been
demonstrated in a recent work \cite{CDG} that if the phase $\Phi_\mu$ is set
to be 0 or $\pi$, the full range of $\Phi_1$ is allowed. In this light, we set
$\Phi_\mu$ to be zero and vary $\Phi_1$ in estimating the CP--odd asymmetry
$A_{\rm CP}$.\s

Figure~\ref{fig:triple_product} shows the $\Phi_1$ dependence of the CP--odd
and CP$\tilde{\rm T}$--even asymmetry $A_{\rm CP}$ of the triple scalar
product.  Numerically we find that the magnitude of the asymmetry can be as
large as 15\% for the parameter set (\ref{eq:parameter_set}), which is
expected to enable us to clearly measure CP violation in the neutralino system
directly with its expected 1--$\sigma$ statistical uncertainty of
$\sqrt{(1-A^2_{\rm CP})/N_{\rm ev}}\simeq 3.1\%$
for the number of events of $N_{\rm ev}=1000$.\s

We note finally that if the measured value of $A_{\rm CP}$ turns to be very
small, close to zero, the two-fold ambiguity $\Phi_1=0$ or $\Phi_1=\pi$, can
be resolved with the help of either the slope parameters $\xi_\pm$ or the
lepton invariant mass/opening angle distribution, see Fig.~\ref{fig:etas} and
Fig.~\ref{fig:mass_angle}, respectively.
%

\section{Conclusions}
\label{sec:conclusion}

Taking into account the possibility of having highly polarized neutralinos
$\tilde{\chi}^0_2$ and
of reconstructing their rest frames, we have performed a systematic analysis
of the polarized neutralino decay $\tilde{\chi}^0_2 \to\tilde{\chi}^0_1
\ell^+\ell^-$  in its decay rest frame
for all the physical implications due to the Majorana nature as
well as CP violation of the neutralino system in the MSSM. 
We have demonstrated that the decay process (\ref{eq:chain})
in the cascade channel like (\ref{cascade})  can 
be used to provide alternative powerful methods for probing the neutralino
system in detail.\s

In the CP invariant case, the Majorana nature of the neutralinos can be checked
through the leptonic decay $\tilde{\chi}^0_2\to\tilde{\chi}^0_1\ell^+\ell^-$
by verifying that in the $\tilde{\chi}^0_2$ rest frame:
\begin{itemize}
\item the charged lepton energy distribution is identical irrespective of the
  electric charge of the lepton,
\item the sum of the negative and positive lepton angle distributions is
  independent of the lepton angles with respect to the neutralino polarization
  vector,
\item the relative CP parity of two neutralinos can be identified by measuring
  the threshold behavior of the invariant mass distribution near the kinematic
  end point and/or the dependence of the decay distribution on the opening
  angle near the angle close to $\pi$ and/or $0$.
\end{itemize}
In the CP non--invariant case, if all the absorptive parts (which are expected
to be usually tiny for such an electroweak decay) are ignored, both the lepton
energy and angle distributions in the decay
$\tilde{\chi}^0_2\to\tilde{\chi}^0_1\ell^+\ell^-$ can allow us to probe the
Majorana nature of the neutralinos as in the CP invariant case.
In addition, the three--body leptonic neutralino 
decays allow us to construct a CP--odd triple scalar product of the neutralino
polarization vector and two lepton momenta. Numerically, we have found that
for the parameter set (\ref{eq:parameter_set}) the CP asymmetry related to the
triple product could be of the order of 15\%, while satisfying the severe EDM
constraints on the SUSY parameters. \s

Finally, we emphasize that the analyses in the present work are only
statistical and based on the assumption that both the degree of neutralino
polarization and the efficiency of reconstructing the neutralino rest frame
are 100\%. Clearly, it is necessary to perform further detailed experimental
analyses with realistic values of reconstruction efficiencies and with
background
processes included, which is beyond the scope of the present theoretical
analyses. However, we think that
the results of our theoretical studies are encouraging enough to
motivate further detailed experimental investigations to assess the
feasibility for
probing the Majorana nature of neutralinos and CP violation in the neutralino
system at future colliders.\s

\section*{Acknowledgments}

The authors thank J.A. Aguilar--Saavedra and H. Baer for their valuable
comments, and Gudrid Moortgat-Pick and Hans-Ulrich Martyn for critical
  remarks 
and reading of the manuscript. 
The work of SYC was supported in part by the Korea Research
Foundation Grant (KRF--2004--041--C00081) and in part by KOSEF through CHEP at
Kyungpook National University and the work of YGK was supported by the Korean
Federation of Science and Technology Societies through the Brain Pool program.
JK and KR are supported by the KBN Grant 2 P03B 040 24 for years 2003-2005 and
115/E-343/SPB/DESY/P-03/DWM517/2003-2005.\s


\end{document}